\documentclass[acmtog, screen, balance = false]{acmart}

\acmSubmissionID{881}
\acmJournal{TOG}
\setcitestyle{square}

\setcitestyle{square}

\copyrightyear{2024}
\acmYear{2024}
\setcopyright{acmlicensed}\acmConference[SIGGRAPH Conference Papers '24]{Special Interest Group on Computer Graphics and Interactive Techniques Conference Conference Papers '24}{July 27-August 1, 2024}{Denver, CO, USA}
\acmBooktitle{Special Interest Group on Computer Graphics and Interactive Techniques Conference Conference Papers '24 (SIGGRAPH Conference Papers '24), July 27-August 1, 2024, Denver, CO, USA}
\acmDOI{10.1145/3641519.3657484}
\acmISBN{979-8-4007-0525-0/24/07}

\usepackage{wrapfig}
\usepackage{nicefrac}
\usepackage{mathtools}
\usepackage{graphicx}
\usepackage{booktabs} 
\usepackage{combelow} 
\usepackage[ruled,vlined,linesnumbered]{algorithm2e}
\usepackage{tabularx} 
\usepackage{colortbl} 
\usepackage{cancel}
\usepackage{makecell} 
\usepackage{enumitem}
\usepackage{manfnt} 
\usepackage[nameinlink]{cleveref}

\usepackage{amsmath}
\usepackage{amsfonts}
\usepackage{amsbsy}

\usepackage{listings} 
\usepackage{courier}
\definecolor{mygreen}{rgb}{0,0.6,0}
\definecolor{mygray}{rgb}{0.5,0.5,0.5}
\definecolor{mymauve}{rgb}{0.58,0,0.82}
\definecolor{superlightgray}{RGB}{240,240,240}
\definecolor{derekTableBlue}{RGB}{189,235,252}

\usepackage[labelfont=bf,format=plain]{caption}
\usepackage{subcaption}
\definecolor{darkBlue}{RGB}{0, 94, 184}
\definecolor{derekBlue}{RGB}{144,210,236}
\definecolor{forest}{RGB}{50,140,90}
\definecolor{derekTableBlue}{RGB}{189,235,252}
\definecolor{iglGreen}{RGB}{153,203,67}
\definecolor{coralRed}{RGB}{250,114,104}
\definecolor{gray}{RGB}{180,180,180}
\definecolor{orange}{RGB}{255,165,0}
\definecolor{TechnionBlue}{RGB}{8,33,78}
\definecolor{Purple}{RGB}{137, 99, 198}
\definecolor{lightgray}{gray}{0.65}

\newcommand{\edit}[1]{#1}

\SetCommentSty{commentFont}
\SetNlSty{textbf}{}{.}

\newcommand*{\refsec}[1]{%
  \begingroup
    \def\sectionautorefname{Sec.}%
    \def\subsectionautorefname{Sec.}%
    \def\subsubsectionautorefname{Sec.}%
    \autoref{sec:#1}%
  \endgroup
}
\newcommand*{\refequ}[1]{%
  \begingroup
    \def\equationautorefname{Eq.}
    \autoref{equ:#1}%
  \endgroup
}

\newcommand*{\reffig}[1]{%
  \begingroup
    \def\figureautorefname{Fig.}%
    \autoref{fig:#1}%
  \endgroup
}

\newcommand*{\reftab}[1]{%
  \begingroup
    \def\tableautorefname{Tab.}%
    \autoref{tab:#1}%
  \endgroup
}
\newcommand*{\refapp}[1]{%
  \begingroup
    \def\appendixautorefname{App.}%
    \autoref{app:#1}%
  \endgroup
}

\newcommand{\R}{\mathbb{R}}

\newcommand{\bool}{\mathcal{B}}

\newcommand{\intersection}{\textsc{intersection}}
\newcommand{\union}{\textsc{union}}
\newcommand{\difference}{\textsc{difference}}
\renewcommand{\complement}{\mathcal{C}}
\renewcommand{\min}{\textit{min}}
\renewcommand{\max}{\textit{max}}

\newcommand{\vecFont}[1]{\mathbf{#1}}

\def\vc{{\vecFont{c}}}

\begin{document}
\title{A Unified Differentiable Boolean Operator with Fuzzy Logic} 

\author{Hsueh-Ti Derek Liu}
\affiliation{%
  \institution{Roblox, University of British Columbia}
  \country{Canada}
  }
\email{hsuehtiliu@roblox.com} 

\author{Maneesh Agrawala}
\affiliation{%
  \institution{Stanford University, Roblox}
  \country{USA}
  }
\email{maneesh@cs.stanford.edu} 

\author{Cem Yuksel}
\affiliation{%
  \institution{University of Utah, Roblox}
  \country{USA}
  }
\email{cem@cemyuksel.com} 

\author{Tim Omernick}
\affiliation{%
  \institution{Roblox}
  \country{USA}
  }
\email{tomernick@roblox.com} 

\author{Vinith Misra}
\affiliation{%
  \institution{Roblox}
  \country{USA}
  }
\email{vmisra@roblox.com} 

\author{Stefano Corazza}
\affiliation{%
  \institution{Roblox}
  \country{USA}
  }
\email{stefano@roblox.com} 

\author{Morgan McGuire}
\affiliation{%
  \institution{Roblox, McGill University}
  \country{Canada}
  }
\email{morgan@roblox.com} 

\author{Victor Zordan}
\affiliation{%
  \institution{Roblox}
  \country{USA}
  }
\email{vbzordan@roblox.com} 
\renewcommand{\shortauthors}{Liu, et al.}









\begin{teaserfigure}
\centering
    \includegraphics[width=\textwidth]{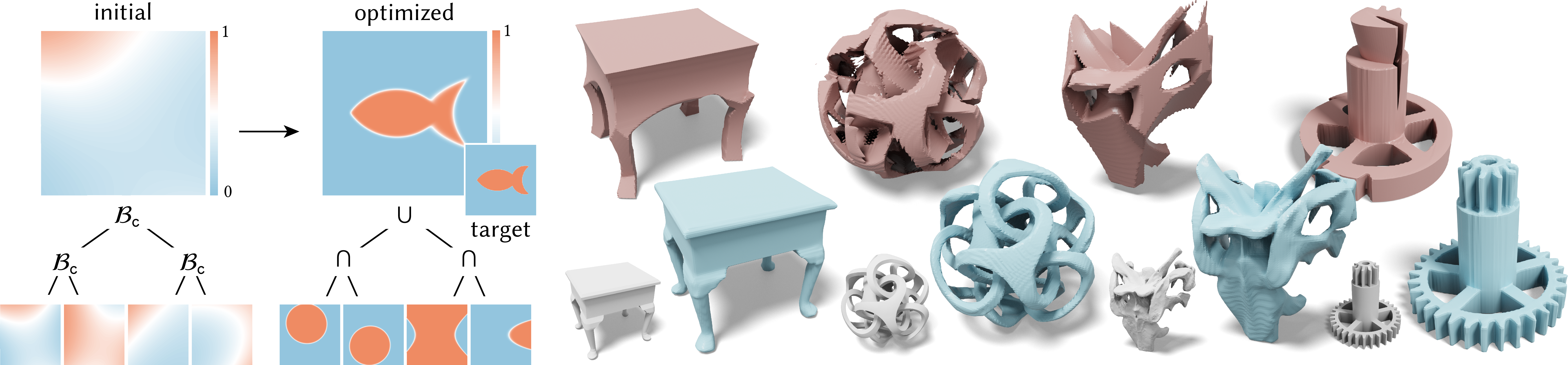}
    \caption{We develop a unified boolean operator $\bool_\vc$ that is differentiable with respect to the type of boolean operations. In the context of inverse CSG, starting with randomly initialize primitives and boolean operations (left tree), our method enables \emph{continuous} optimization on both the primitives and the boolean operations in order to fit the target shape (middle tree). Performing inverse CSG fitting to the ground truth shape (grey) with our method leads to significant quality improvement (blue) over the traditional boolean operations with the \min{} and \max{} operators (red).}
    \label{fig:teaser}
\end{teaserfigure}

\begin{abstract}
This paper presents a unified differentiable boolean operator for implicit solid shape modeling using Constructive Solid Geometry (CSG). Traditional CSG relies on \min, \max{} operators to perform boolean operations on implicit shapes. But because these boolean operators are discontinuous and discrete in the choice of operations, this makes optimization over the CSG representation challenging. Drawing inspiration from \emph{fuzzy logic}, we present a unified boolean operator that outputs a continuous function and is differentiable with respect to operator types. This enables optimization of both the primitives and the boolean operations employed in CSG with continuous optimization techniques, such as gradient descent. We further demonstrate that such a continuous boolean operator allows the modeling of both sharp mechanical objects and smooth organic shapes with the same framework. Our proposed boolean operator opens up new possibilities for future research toward fully continuous CSG optimization.
\end{abstract}

\maketitle

\section{Introduction}
\begin{figure}
    \begin{center}
    \includegraphics[width=1\linewidth]{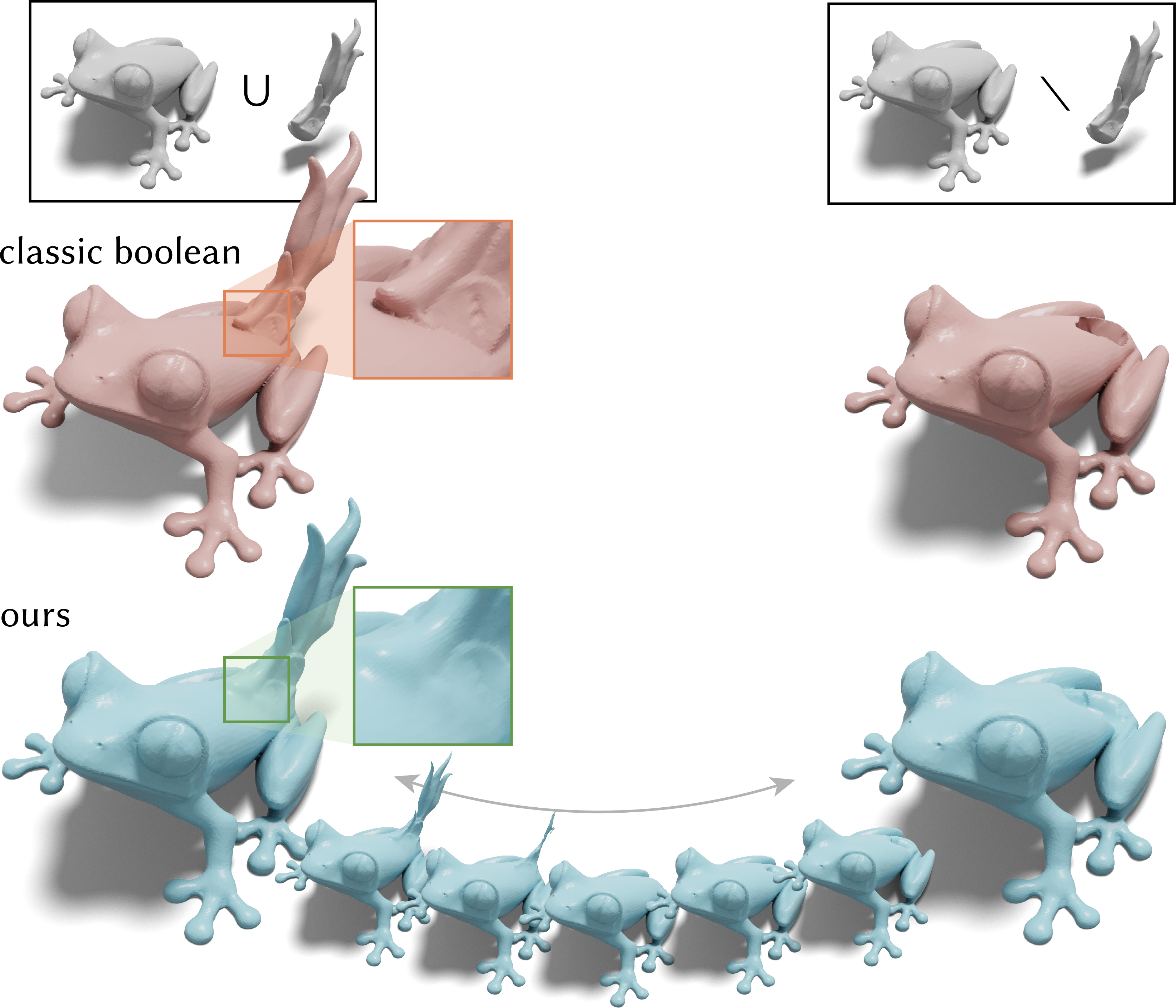}
    \end{center}
    \caption{We present a differentiable boolean operator with respect to its operands and the operator. Given two implicit shapes, our boolean operator can control the blend region between two shapes and outputs a smoothly differentiable function (see the zoom-in part). One can also continuously switch the operator from one to another, such as from {\union} (left) to {\difference} (right).}
    \label{fig:unified_operator}
\end{figure}

Boolean operations are a central ingredient in Constructive Solid Geometry (CSG) -- a modeling paradigm that represents a complex shape using a collection of primitive shapes which are combined together via boolean operations (\intersection, \union, and \difference). CSG provides a precise, hierarchical representation of solid shapes and is widely used in \edit{computer graphics}. 

The importance of CSG has motivated researchers to investigate the \emph{inverse} problem; constructing a CSG tree for a given 3D model from a collection of parameterized primitive shapes.  
A common approach is to treat this as an optimization problem that involves choosing the structure of the CSG tree; the type of boolean operation to perform at each internal node in the tree, as well as the parameters and type (e.g., sphere, cube, cylinder) of the leaf node primitive shapes.
The optimization is difficult because it contains a mixture of discrete (type of boolean operation, number and type of primitive shapes) and continuous (parameters of primitives e.g., radius, width, etc.) variables. Moreover, the degrees of freedom grow exponentially with the complexity of the CSG tree, making the optimization landscape very challenging to navigate.

Previous attempts either tackle the inverse optimization directly with \emph{evolutionary algorithms} \cite{FriedrichFGL19}, or relax some of the discrete variables into continuous variables to reduce the discrete search space.
For instance, one of the discrete decisions is to determine which primitive types (e.g., sphere, cube, cylinder) to use, and a common relaxation is to optimize over a continuously parameterized family of primitives, such as \emph{quadric surfaces} \cite{yu2022capri, dualcsg}. 
This approach allows continuous optimization (e.g., gradient descent) over choosing the type of each primitive, but not the entire tree; the choice of boolean operations and the number of primitives remain discrete variables. 
As a result, these inverse CSG methods pre-determine the structure of the tree including both the boolean operations and the number of primitives and focus on optimizing the primitive parameters. 

In this work, we develop a unified differentiable boolean operator and show how this operator can be used to further relax inverse CSG optimization by turning the discrete choice of boolean operation for each internal CSG node into a continuous optimization variable. 
Drawing inspiration from \emph{Fuzzy Logic}\,\cite{zadeh1965fuzzy}, we first demonstrate how these individual fuzzy logic operations (t-norms, t-conorms) can be applied to boolean operations on solid shapes represented as soft occupancy functions.
Fuzzy boolean operators guarantee that the result remains a soft occupancy function, unlike existing boolean operators (with \min/\max) that operate on signed distance functions.  
These fuzzy booleans on the soft occupancy naturally generalize CSG from modeling shapes with sharp edges to modeling smooth organic shapes. 
We then construct a unified fuzzy boolean operator that uses tetrahedral barycentric interpolation to combine the individual fuzzy boolean operations (see \reffig{unified_operator}). 
We show that our unified operator is differentiable, avoids vanishing gradients and is monotonic making it especially well-suited for gradient-based optimization. 
We apply our unified boolean operator in the context of inverse CSG optimization and find significant improvements in the accuracy of the resulting tree compared to previous methods (see \reffig{teaser}).

\section{Related Work}
Our contribution uses fuzzy logic to design a unified differentiable boolean operator with applications in gradient-based inverse CSG optimization. 
While researchers have formulated inverse CSG as a program synthesis \cite{SharmaGLKM18, WuXZ21, du2018inversecsg}, combinatorial optimization \cite{WuXW18}, or a global optimization \cite{HamzaS04, FriedrichFGL19} problem, we reformulate it as differentiable gradient descent optimization problem with respect to boolean and primitive parameters.
Here, we consider how our work relates to differentiable CSG optimization and to other boolean operators used in geometric modeling. 

\subsection{Gradient-Based CSG Optimization}
Optimizing CSG trees requires determining the structure of the tree, the boolean operations, and the primitive parameters. 
In order to deploy continuous optimization techniques (such as gradient descent), existing solutions rely on pre-defining all the discrete variables (the tree structure and the boolean operations), and then only optimizing the primitive parameters. 
The simplest pre-defined ``CSG tree'' is a union of many parts, including convex shapes \cite{deng2020cvxnet, ChenTZ20} and neural implicit functions \cite{dengunsupervised}.
Some works have also explored more complicated pre-defined tree structures with a mixture of \intersection, \union, \difference{} operators \cite{RenZ0LJCZPZZY21, yu2022capri, dualcsg}.
Instead of pre-determining the boolean operators, \citet{kania2020ucsg} proposed a brute force approach to all possible boolean combinations. However this approach suffers from scalability issues as the number of combinations grows exponentially with respect to the depth of the CSG tree. 

Our contribution complements these techniques by introducing a unified boolean operator which enables continuous optimization on the choice of boolean operations. This avoids the need for brute force or pre-determining boolean operations, leading to better reconstruction (\reffig{teaser}).

\subsection{Boolean Operators in Geometric Modeling}
\begin{figure}
    \begin{center}
    \includegraphics[width=1\linewidth]{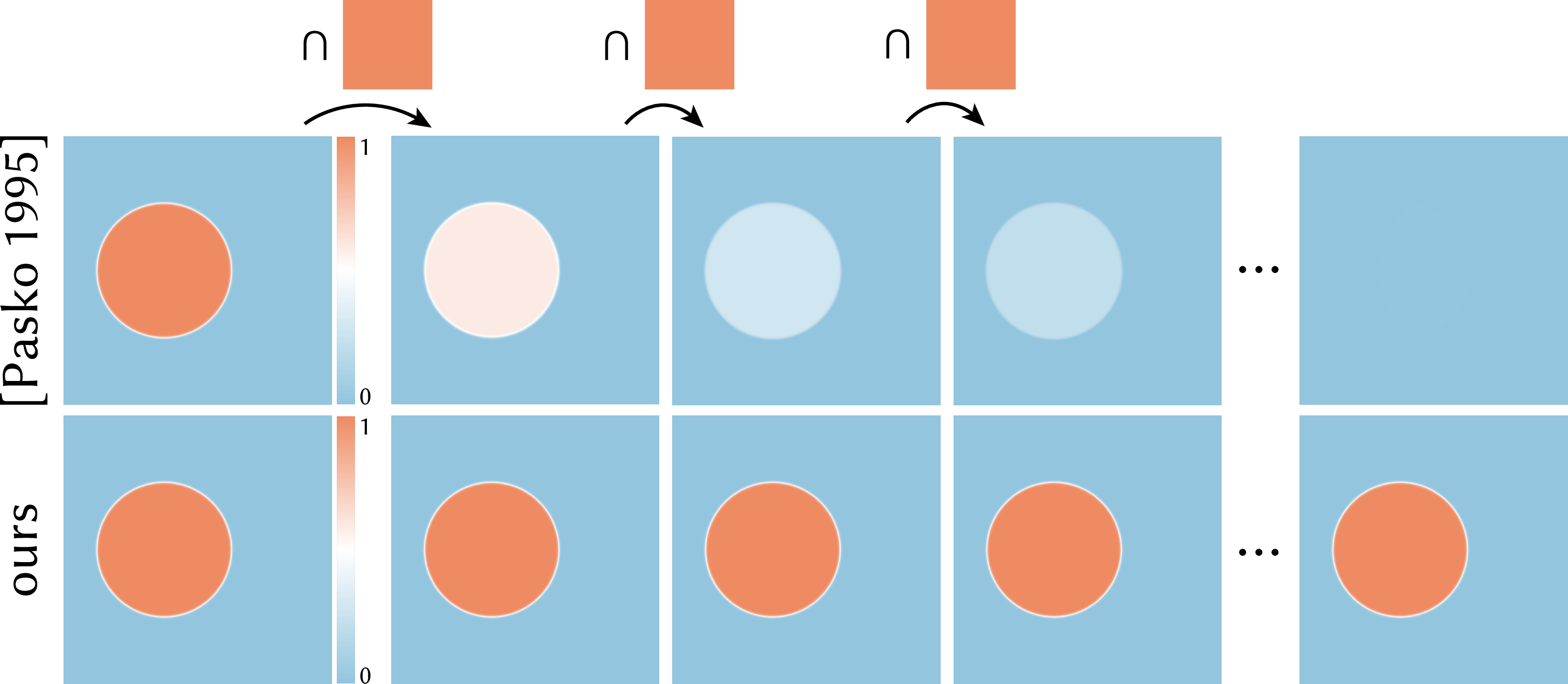}
    \end{center}
    \caption{We show a comparison of our method against a soft boolean operator, R-function, used in \cite{PaskoASS95}. The R-function does not satisfy the axioms characterizing the behavior of \intersection, it produces outputs that do not align with the classic (crisp) \intersection. We show that the R-function (top row) produces a nearly empty shape when intersecting with the full shape multiple times. In contrast, our method matches the behavior of a crisp \intersection. }
    \label{fig:r_function}
\end{figure}
The importance of boolean operations has stimulated research topics in \emph{differentiable} boolean operators in geometric modeling. Despite having the same name, the term ``differentiable boolean'' can refer to (1) boolean operators that output a differentiable function, and (2) unified boolean operators that can be differentiated with respect to the type of operations.

The most common usage of ``differentiable boolean'' refers to boolean operations that produce \emph{differentiable} functions, also known as \emph{soft blending}. 
Traditionally, $\min$ and $\max$ operators are used to produce \intersection{} and \union{} between two implicit functions. But the caveat is that their gradients are ill-defined at locations when the input functions have the same value. 
This motivated \citet{Ricci73} to introduce a soft blending operator using a variant of \emph{p-norms} $\|x\|_p = (\sum_i |x|^{p})^{\nicefrac{1}{p}}$ to produce smooth and differentiable outputs. 
Later on, \citet{PaskoASS95} demonstrated the use of the \emph{Rvachev function} (R-function) \cite{Rfunction} to define a soft boolean operation that outputs $C^n$ continuous implicit functions with user controllable $n$. 
Several works \cite{WyvillGG99, BartheWG04, gourmel2013gradient} further extended this soft boolean formulation to provide fine controls on the blended region. 
Alternative formulations based on polynomial smoothing \cite{InigoSmoothMin} or fuzzy logic \cite{fuzzy_blend} are viable choices as well.
These approaches enable continuous outputs of individual boolean operations, but, in contrast to ours, they did not focus on differentiating through different boolean operations.

Another usage of ``differentiable boolean'' refers to unified boolean operators that can differentiably switch from one operation to another. \citet{WyvillGG99} show that a modified R-function is an instance of a unified boolean operator with a smooth transition between \union{} and \intersection{}. 
However, we demonstrate that the R-function does not satisfy important axioms of boolean operators (see \refsec{fuzzy_logic}), specifically the \emph{boundary condition}. 
This implies that the R-function is prone to unexpected behavior. For instance, in \reffig{r_function} we show that if we intersect a shape with the full shape multiple times using the R-function, it ends up producing an almost-empty shape. 
Our method, instead, possesses the properties of a \emph{valid} boolean operator (see \refsec{fuzzy_logic}) and guarantees to match the expected behavior of classic boolean operations when the inputs are binary. 

Our proposed boolean operator is relevant to both usages of ``differentiable boolean''; our method outputs continuous functions and can be differentiated through different boolean types. 
We demonstrate applications in modeling smooth shapes and inverse CSG optimization in \refsec{applications}.

\section{Background} \label{sec:background_fuzzy_logic}
The concept of fuzzy logic \cite{zadeh1965fuzzy} has applications in a wide variety of problem domains \cite{dzitac2017fuzzy}. \edit{In computer graphics,} fuzzy logic has been used in image processing \cite{chacon2006fuzzy}, color compositing \cite{smith1995image}, and spline interpolation \cite{fuzzy_blend}. Here we summarize the core ideas of fuzzy logic. 

\subsection{Fuzzy Set}\label{sec:fuzzy_set}
A fuzzy set $X = (P, f_X)$ is a tuple of the universe of elements $P = \{ p \}$ and a \emph{membership} function  $f_X: P \rightarrow [0, 1]$ such that $f_X(p) = 0$ implies that element $p$ is not a member of $X$, $f_X(p) = 1$ implies $p$ is a full member of $X$ and $0 < f_X(p) < 1$ implies $p$ is a partial member of $X$. 
Fuzzy sets are a generalization of the classic ``crisp'' set, whose membership function only outputs 0 or 1.
For instance, suppose we define a fuzzy set $H = (P, f_H)$ of the temperatures one considers hot. The universe of elements $P$ are all possible temperature values. One might consider some temperature values, such as 40 degrees Celsius, as full members of $H$ so that $f_H(40^\circ C) = 1$. But 25 degrees Celsius, might only be a partial member $f_H(25^\circ H) = 0.3$. Fuzzy sets model this notion of partial membership.

\subsection{Fuzzy Logic}\label{sec:fuzzy_logic}
Fuzzy logic develops boolean operations on fuzzy sets. 
Given two fuzzy sets $X = (P, f_X)$ and $Y = (P, f_Y)$, a boolean operation is defined on the membership function. For instance, \intersection{} $\cap$, \union{} $\cup$, and \textsc{complement} $\neg$ between fuzzy sets are defined as
\begin{align}
    X \cap Y = (P, f_{X \cap Y}), \quad X \cup Y = (P, f_{X \cup Y}), \quad \neg X = (P, f_{\neg X}).
\end{align}
A core question in fuzzy logic research is how to define these boolean membership functions $f_{X \cap Y},  f_{X \cup Y}, f_{\neg X}$. A very common approach is to define them using the $\min, \max$ operators\,\cite{Godel1932} as
\begin{align}
    &f_{X \cap Y} = \min \big( f_X(p), f_Y(p) \big) = \min(x, y),\\
    &f_{X \cup Y} = \max \big( f_X(p), f_Y(p) \big) = \max(x, y),\\
    &f_{\neg X} = 1 - f_X(p) = 1 - x.
\end{align}
To simplify notation, here and for the rest of the paper, we use the lowercase letter $x$ to refer to $f_X(p)$, the membership function of $X$ applied to a generic element $p \in P$. Similarly, $y$ refers to $f_Y(p)$. 

\vspace{0.5em}
\noindent 
{\bf \em Fuzzy Intersection.}
Fuzzy logic researchers have explored other definitions of $f_{X \cap Y}$ \cite{fuzzy_logic_book}. Suppose ${f_{X \cap Y} = \top(x, y)}$.
They define $\top$ as a valid intersection function when the following axioms hold:
\begin{align}
    &\top(x, 1) = x  \tag{boundary condition}\\
    &\top(x, y) \leq \top(x, z), \text{if } y \leq z  \tag{monotonicity}\\
    &\top(x, y) = \top(y, x)  \tag{commutativity}\\
    &\top(x, \top(y, z)) = \top(\top(x, y), z)  \tag{associativity}
\end{align}
where $x = f_X(p), y = f_Y(p), z = f_Z(p) \in [0, 1]$ are fuzzy membership values for generic element $p$.
Any function that satisfies these axioms is called a \emph{t-norm} $\top$ \cite{Menger1942}. 
These axioms ensure that the behavior of the fuzzy intersection operator converges to the classic intersection (\textsc{AND}) operator when membership values are binary.
Some popular t-norms include G\"{o}del's\,\shortcite{Godel1932} minimum where $\top(x,y)=\min(x,y)$, product $\top(x,y) = x \cdot y$, {\L}ukasiewicz $\top(x,y)=\max(0, x+y-1)$, and Yager\,\shortcite{yager1980general} $\top(x,y)=\max(1 - ((1-x)^p + (1-b)^p)^{1/p}, 0)$. 

\vspace{0.5em}
\noindent 
{\bf \em Fuzzy Union.}
Similarly, suppose $f_{X \cup Y} = \bot(x, y)$. In fuzzy logic, $\bot$ is a valid \union{} function if:
\begin{align}
    &\bot(x, 0) = x  \tag{boundary condition}\\
    &\bot(x, y) \leq \bot(x, z), \text{if } y \leq z \tag{monotonicity}\\
    &\bot(x, y) = \bot(y, x) \tag{commutativity}\\
    &\bot(x, \bot(y, z)) = \bot(\bot(x, y), z) \tag{associativity}
\end{align}
The functions that satisfy these axioms are known as \emph{t-conorms} $\bot$. 
Common t-conorms include G\"{o}del's\,\shortcite{Godel1932} maximum $\bot(x,y)=\max(x,y)$, probabilistic sum $\bot(x,y)=x + y - x\cdot y$, bounded sum $\bot(x,y)=\min(x+y, 1)$, and Yager\,\shortcite{yager1980general} $\bot(x,y)=\min( (x^p + y^p)^{1/p}, 1 )$.

\vspace{0.5em}
\noindent 
{\bf \em Fuzzy Complement.}
The set of axioms for defining a valid \textsc{complement} function $f_{\neg X} = \complement (x)$ are
\begin{align}
    &\complement (0) = 1, \complement (1) = 0  \tag{boundary condition}\\
    &\text{if } x \leq y \text{, then } \complement (x) < \complement (y) \tag{monotonicity}
\end{align}
Valid complement functions include cosine $\complement(x) = \nicefrac{1 + \cos(\pi x)}{2}$, Sugeno $\complement(x) = \nicefrac{1-x}{1 + \lambda x}$ with $\lambda \in (-1, \infty)$, and Yager\,\shortcite{yager1980general} $\complement(x) = (1 - x^\lambda)^{1/\lambda}$. The widely used complement $\complement(x) = 1-x$, is the Yager complement with $\lambda = 1$.

\begin{wrapfigure}[7]{r}{1.4in}
    \vspace{-12pt}
    \includegraphics[width=\linewidth, trim={50mm 0mm 0mm 0mm}]{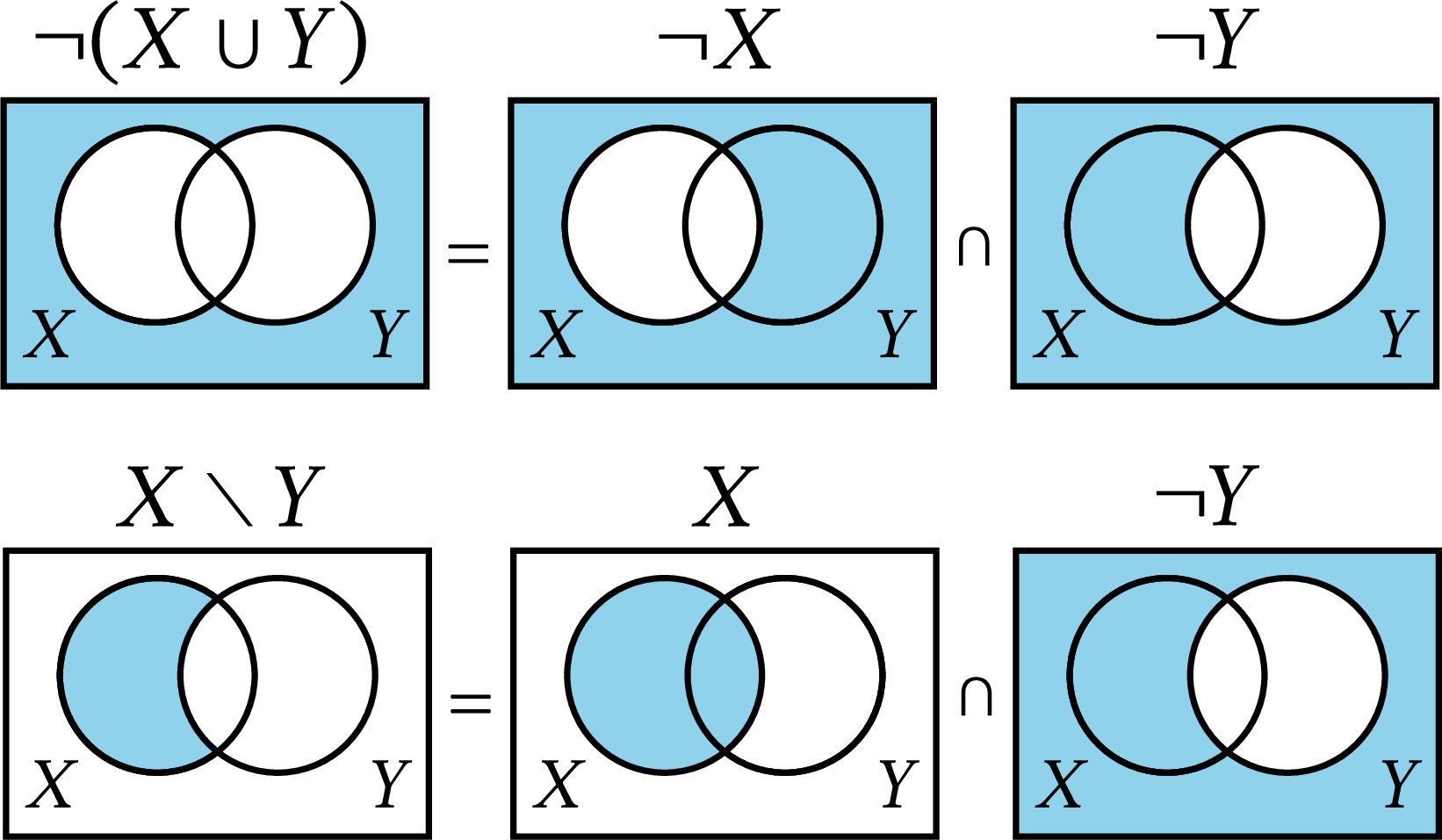}
\end{wrapfigure}
\vspace{0.5em}
\noindent
{\bf \em Fuzzy Difference.}
In fuzzy logic, the \difference{} operator $\setminus$ is usually derived from the De Morgan's laws (see inset), which state the relationship between the \union, \intersection, and \textsc{complement} operators,
\begin{align}\label{equ:duality}
   \neg (X \cup Y) =  \neg X \cap \neg Y,
\end{align}
and the relationship between \difference{} and the other operators
\begin{align}\label{equ:de_morgan}
    X \setminus Y = X \cap \neg Y.
\end{align}
For the De Morgan's law to hold, one has to jointly define the \intersection, \union, and \textsc{complement} operators so that they satisfy \refequ{duality}. 
Then a valid {\difference} operator $\setminus$ can be derived 
from {\intersection} and \textsc{complement} using \refequ{de_morgan} as 
\begin{align}
    f_{X \setminus Y} = \top(x, \complement (y)).
\end{align}

\section{A Unified Differentiable Boolean Operator} 
\label{sec:boolean}
To apply fuzzy logic to CSG modeling, we interpret a solid shape, represented by a soft occupancy function, as a fuzzy set $X = \{P, f_X\}$. Here, $P = \{ p\}$ denotes the universe of points $p \in \R^d$ and the membership function $f_X: P \rightarrow [0, 1]$ is the soft occupancy function representing the probability of a point lying inside the shape. 
Then we can directly apply the fuzzy boolean operations presented in \refsec{background_fuzzy_logic}. 
However, we must choose \intersection, \union, and \textsc{complement} appropriate to our task. We first present our choice for each of these functions (\refsec{product_fuzzy_logic}) and then describe how to combine them into a unified boolean operator (\refsec{unified_operator}).

\subsection{Product Fuzzy Logic}\label{sec:product_fuzzy_logic}
Motivated by our goal of continuous optimization, we would like each of our individual fuzzy boolean operations {\intersection} $\top$, {\union} $\bot$ and \textsc{complement} $\complement$ to be differentiable and have non-vanishing (i.e. non-zero) gradients with respect to their inputs. Vanishing gradients can result in plateaus in the energy landscape making gradient-based optimization difficult. 

Boolean operators as defined by the product fuzzy logic meet these criteria. Specifically they are defined as
\begin{align} \label{equ:product_fuzzy_logic}
    &f_{X \cap Y} = \top(x,y) = xy \\
    &f_{X \cup Y} = \bot(x,y) = x + y - xy \\
    &f_{\neg X} = \complement(x) = 1 - x 
\end{align}
where $X$ and $Y$ are two solid shapes and $x = f_X(p), y = f_Y(p) \in [0, 1]$ are their soft occupancy values at a generic point $p$. 
These definitions satisfy the axioms of valid boolean operators (see \refsec{background_fuzzy_logic}). They correspond to valid t-norm $\top$, t-conorm $\bot$, and complement $\complement$ functions, respectively, in fuzzy logic. 
They also satisfy De Morgan's law \refequ{duality}, allowing us to compute {\difference} as
\begin{align}
   f_{X \setminus Y} = x - xy, \quad f_{Y \setminus X} = y - xy
\end{align}

\begin{wrapfigure}[9]{r}{1.4in}
    \vspace{-12pt}
	\includegraphics[width=\linewidth, trim={50mm 0mm 0mm 0mm}]{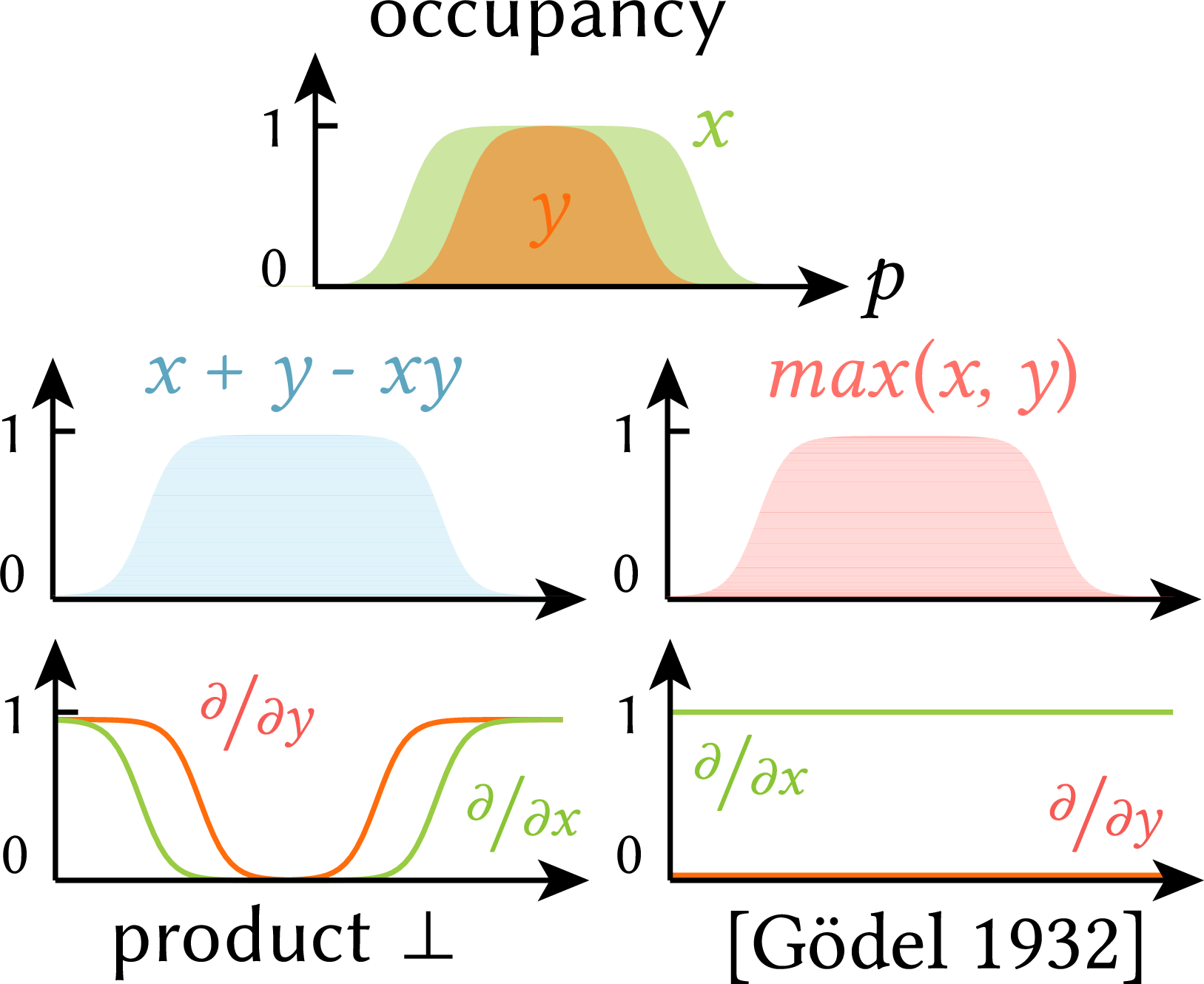}
	\label{fig:vanishing_grad_1D}
\end{wrapfigure}
\begin{figure}
    \begin{center}
    \includegraphics[width=1\linewidth]{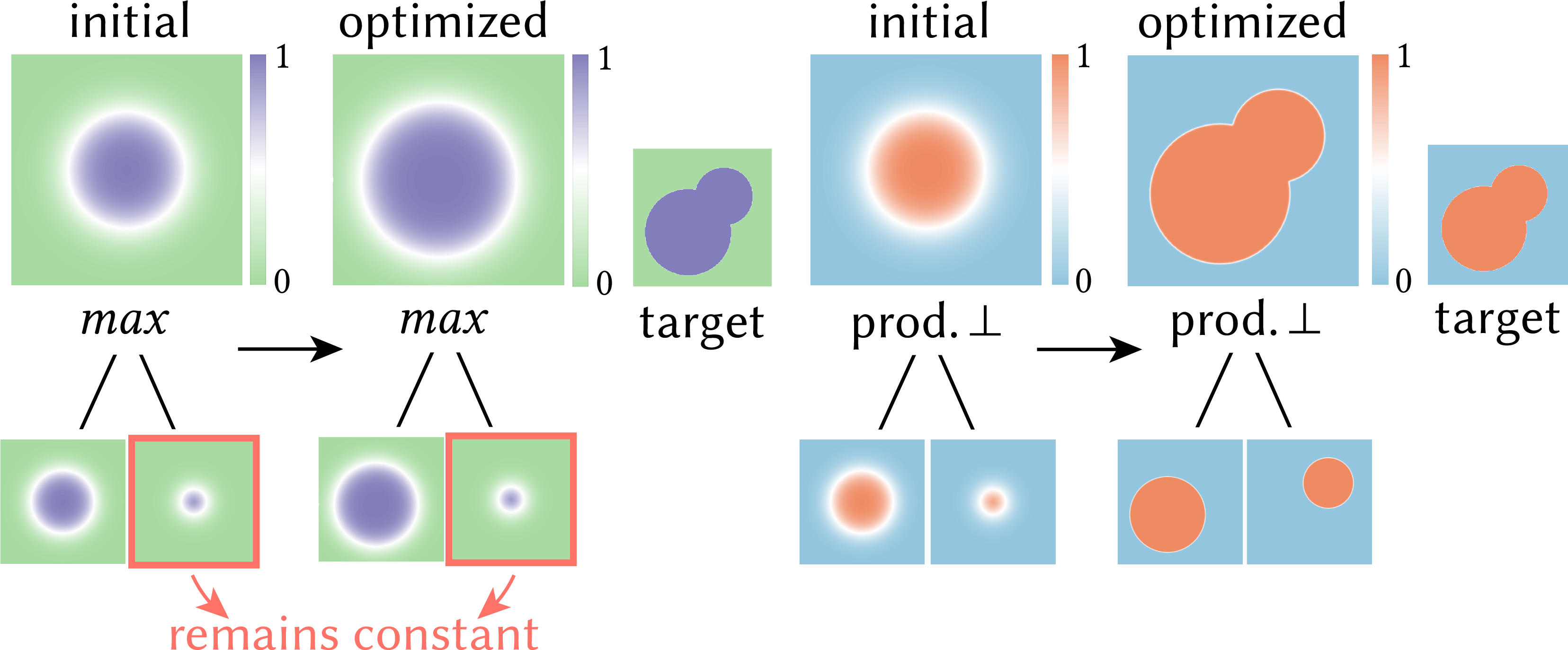}
    \end{center}
    \caption{We demonstrate the importance of avoiding vanishing gradients in an inverse CSG for fitting the union of two circles. 
    Using the traditional \max{}\cite{Godel1932} operator suffers from vanishing gradients, leading to a primitive (left, red) remaining unchanged throughout the optimization and thus failing to reconstruct the target shape. 
    In contrast, our method presented in \refequ{product_fuzzy_logic} avoids vanishing gradient and is able to recover the ground truth (right).}
    \label{fig:vanishing_grad}
\end{figure}

The product fuzzy logic boolean functions are differentiable with respect to their inputs $x$ and $y$. 
Other fuzzy logic functions, such as G\"{o}del's\,\shortcite{Godel1932} \min/\max, t-norm/t-conorm, are not differentiable at singularities.
In addition, the product fuzzy logic functions are also much less prone to {\em vanishing gradients} compared to many other fuzzy logic function definitions\,\cite{van2022analyzing}.
More formally vanishing gradients occur when the partial derivatives $\nicefrac{\partial}{\partial x}, \nicefrac{\partial}{\partial y}$ equal zero (or become very small). 
In the inset, we illustrate a 1D example where occupancy values $x$ are strictly larger than or equal to $y$. Defining \union{} with the G\"{o}del's \max{} operator results in a zero gradient for $y$, as $\nicefrac{\partial}{\partial y} = 0$. In contrast, using the \union{} defined in \refequ{product_fuzzy_logic} still possesses non-zero gradients for both $x, y$.
In \reffig{vanishing_grad}, we further demonstrate the importance of avoiding vanishing gradient in a simple example of our inverse CSG task with continuous optimization (see \refsec{inverse_csg_GD} for implementation details). 

\subsection{Unifying Boolean Operations}\label{sec:unified_operator}
To create a unified fuzzy boolean operator that is differentiable with respect to the type of boolean operation ({\intersection}, {\union}, {\difference}), our approach is to interpolate their respective membership functions using a set of interpolation control parameters $\vc$.
Our goal is to design an interpolation scheme that is continuous and monotonic in the parameters $\vc$ so that the interpolation function avoids unnecessary local minima.

\begin{figure}
    \begin{center}
    \includegraphics[width=1\linewidth]{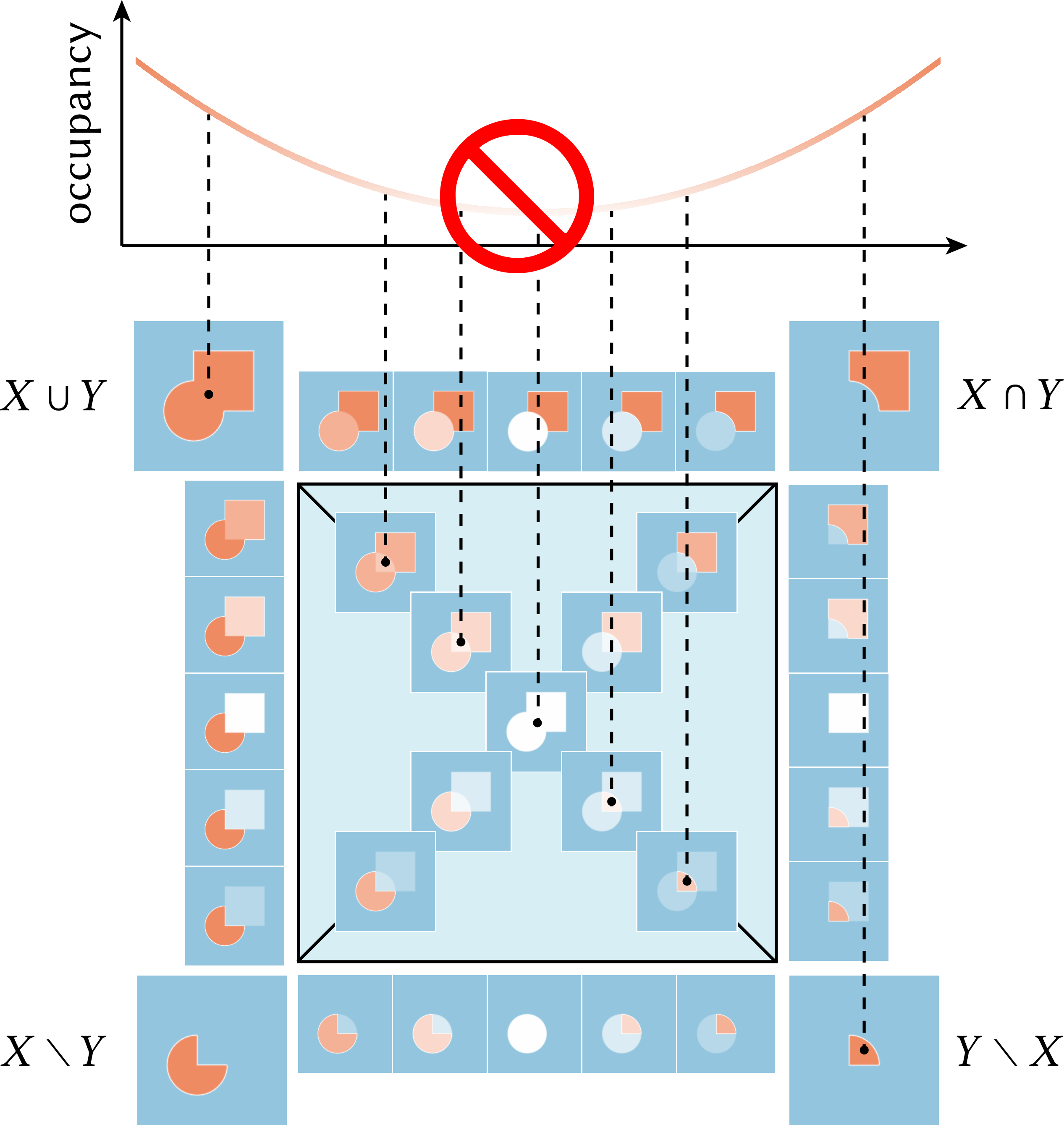}
    \end{center}
    \caption{Naive bilinear interpolation introduces additional local minima in the occupancy value (top plot) when interpolating between, for instance, \union (top left) and \textsc{difference} (bottom right).}
    \label{fig:bilinear_interpolation}
\end{figure}

A naive solution is to use bilinear interpolation between the four boolean operations $f_{X \cap Y}, f_{X \cup Y}, f_{X \setminus Y}, f_{Y \setminus X}$. While such interpolation can look smooth, bilinear interpolation exhibits non-monotonic changes and creates local minima in the interpolated occupancy (see \reffig{bilinear_interpolation}). 
This is because bilinear interpolation implicitly forces the average between $f_{X \cup Y}, f_{Y \setminus X}$ and the average between $f_{X \cap Y}, f_{X \setminus Y}$ to be equivalent. In many cases, these averages are not equivalent and thus the constraint forces the interpolation to be non-monotonic.

\begin{figure}
    \begin{center}
    \includegraphics[width=1\linewidth]{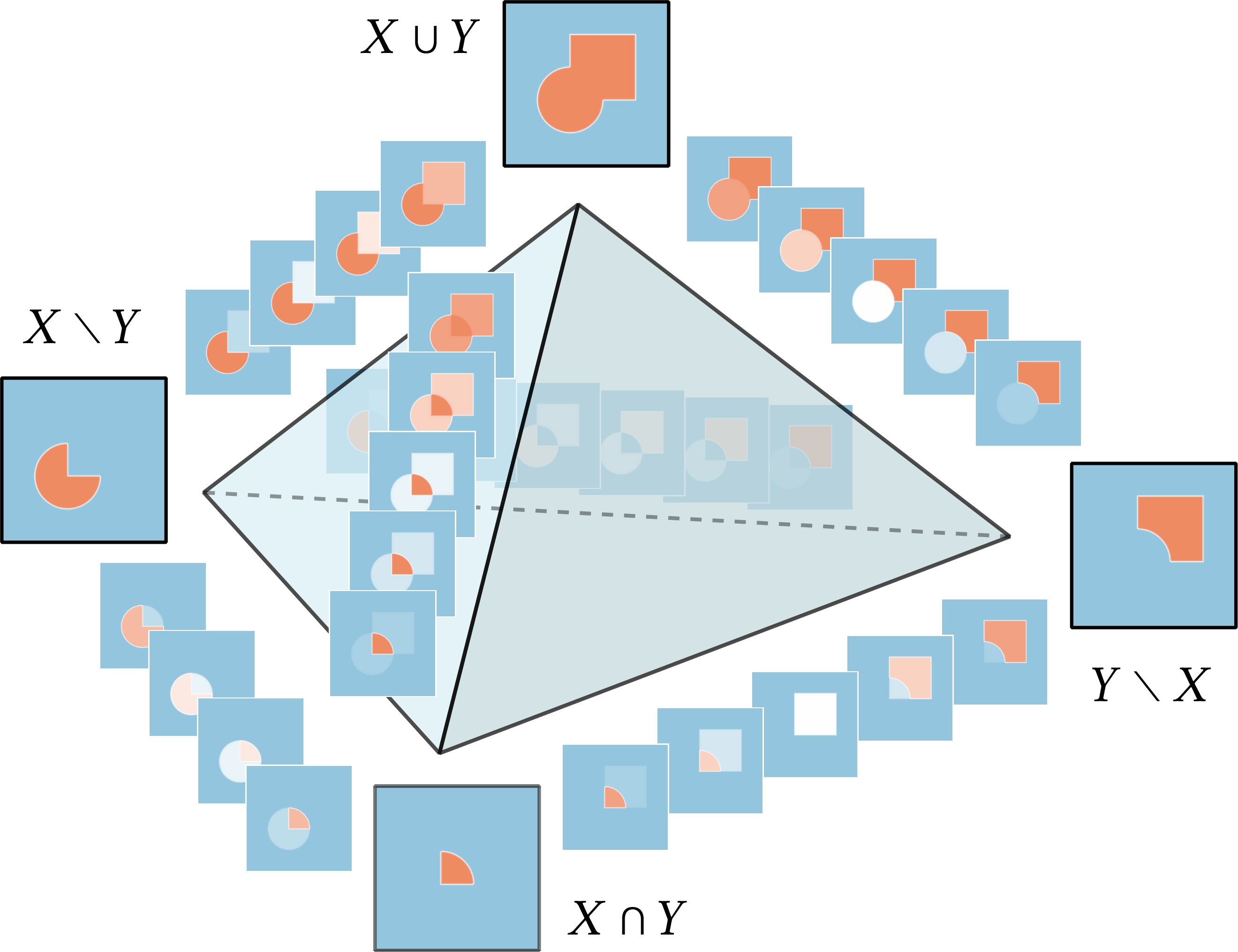}
    \end{center}
    \caption{We use barycentric interpolation to produce monotonic interpolation between different boolean operators. This avoids undesired local minima in occupancy interpolation.}
    \label{fig:boolean_tetrahedron}
\end{figure}

Instead of this, we use tetrahedral barycentric interpolation. More specifically we treat individual boolean operations (\union, \intersection, and two \difference{s}) as vertices of a tetrahedron and define our unified boolean operator function $\bool_\vc$ as barycentric interpolation within it as
\begin{align}\label{equ:unified_boolean}
    \boxed{\bool_\vc (x, y) = (c_1 + c_2) x + (c_1 + c_3) y + (c_0 - c_1 - c_2 - c_3) xy }
\end{align} 
where $\vc = \{ c_0, c_1, c_2, c_3 \}$ are parameters that control the type of boolean operations and they satisfy the properties of barycentric coordinates
\begin{align}
    0 \leq c_i \leq 1 &&\text{and}&& c_0 + c_1 + c_2 + c_3 = 1 \;.
\end{align} 
When the parameter $\vc$ is a one-hot vector, i.e. the barycentric coordinates for the vertices of a tetrahedron, it exactly reproduces the product logic operators
\begin{alignat}{3} \label{equ:satisfy_product_fuzzy_logic}
    &\bool_{1,0,0,0}(x, y)\ &&= xy\ &&= f_{X \cap Y}  \\
    &\bool_{0,1,0,0}(x, y)\ &&= x + y - xy\ &&= f_{X \cup Y} \\
    &\bool_{0,0,1,0}(x, y)\ &&=  x - xy\ &&= f_{X \setminus Y}  \\
    &\bool_{0,0,0,1}(x, y)\ &&= y - xy\ &&= f_{Y \setminus X}
\end{alignat}
\begin{wrapfigure}[6]{r}{1.4in}
    \vspace{-12pt}
    \includegraphics[width=\linewidth, trim={50mm 0mm 0mm 0mm}]{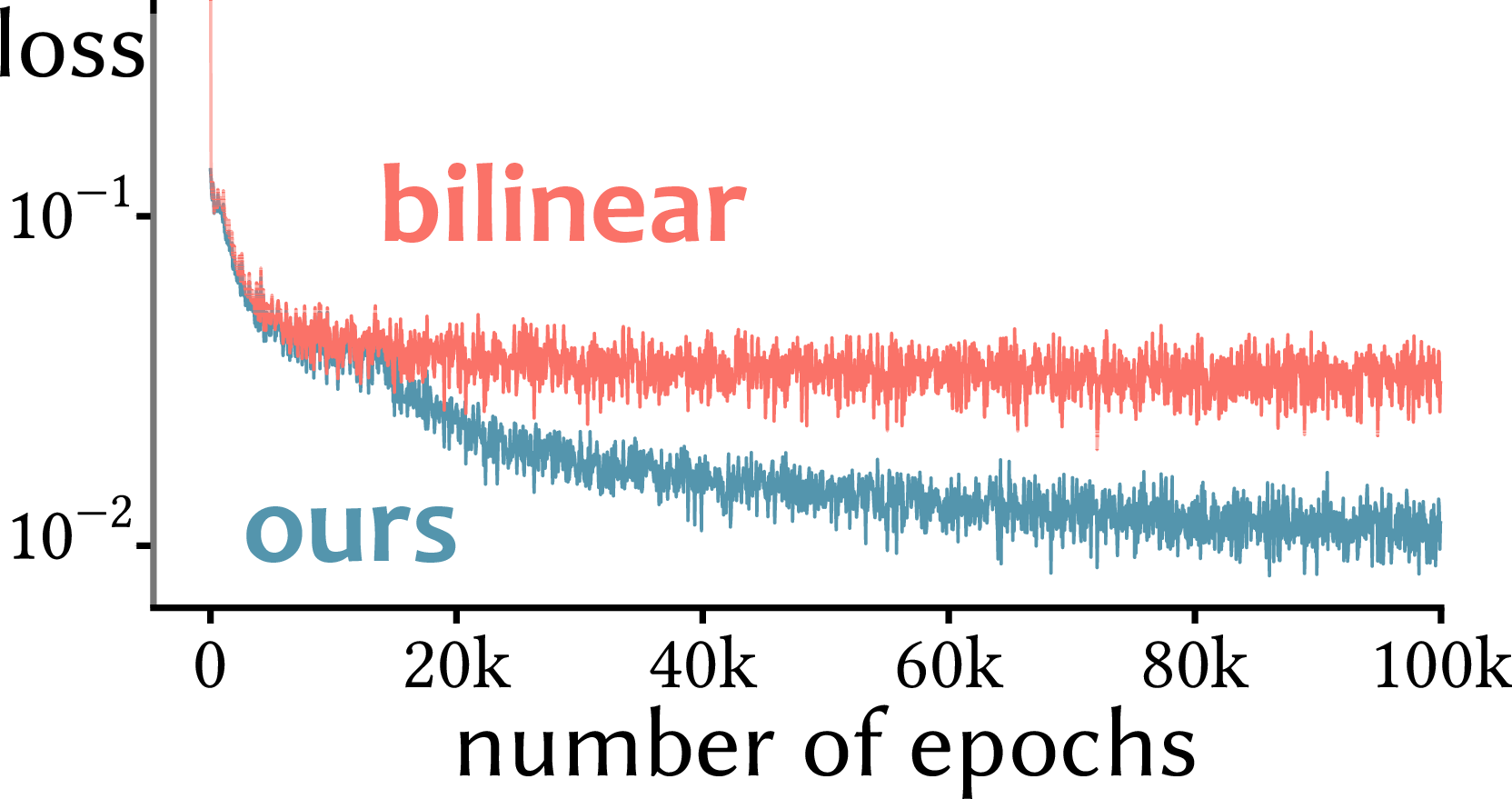}
\end{wrapfigure}
From \refequ{unified_boolean}, we can immediately observe that our unified operator is continuously differentiable with respect to both the inputs $\nicefrac{\partial \bool_\vc}{\partial x}, \nicefrac{\partial \bool_\vc}{\partial y}$ and the control parameters $\nicefrac{\partial \bool_\vc}{\partial c_i}$ by design. 
Moreover, our operator $\bool_\vc$ provides monotonic interpolation between the individual boolean operations at the vertices because interpolation along the edge of a tetrahedron is equivalent to a 1D convex combination (\reffig{boolean_tetrahedron}). Empirically, using barycentric interpolation leads to a smaller error compared to using bilinear interpolation \edit{(see inset)}.
\section{Results}\label{sec:applications}
Building on top of fuzzy logic, we first demonstrate our choice of individual operators from \refequ{product_fuzzy_logic} leads to a natural generalization \edit{from} modeling sharp solid objects to smooth organic shapes in \refsec{fuzzy_csg_system}.
When combined with our unified boolean operator (see \refequ{unified_boolean}), they lead to significant improvements in the inverse CSG tasks, including fitting a single shape \refsec{inverse_csg_GD} or a collection of shapes \refsec{csg_generation}.

\subsection{Fuzzy CSG System}\label{sec:fuzzy_csg_system}
Using fuzzy boolean operators in CSG gives the ability to model both mechanical objects with crisp edges and smooth organic shapes with the same framework. Specifically, if the underlying implicit shapes are crisp binary occupancy functions, our method produces the same sharp results as the traditional CSG. If the input shapes are soft occupancy functions, our method outputs smooth shapes based on the ``softness'' present in the input shape. 

\begin{figure}
    \begin{center}
    \includegraphics[width=1\linewidth]{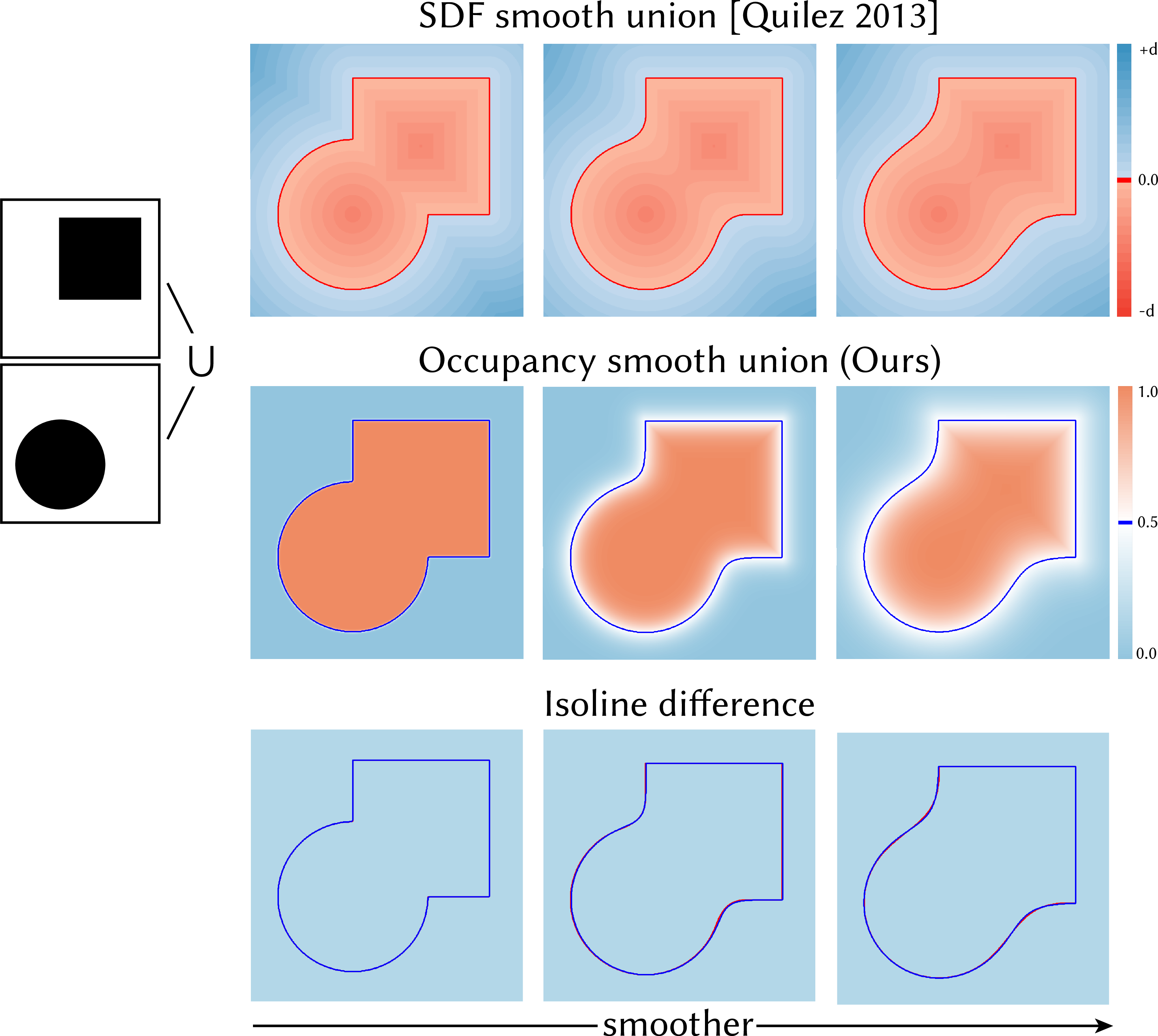}
    \end{center}
    \caption{The soft boolean operator of \citet{InigoSmoothMin} has been demonstrated to be an effective way to model smooth shapes (top row). Our fuzzy boolean operator can also produce smooth boolean results (middle row) with visually indistinguishable isolines (bottom row).}
    \label{fig:soft_boolean}
\end{figure}

\begin{figure}
    \begin{center}
    \includegraphics[width=1\linewidth]{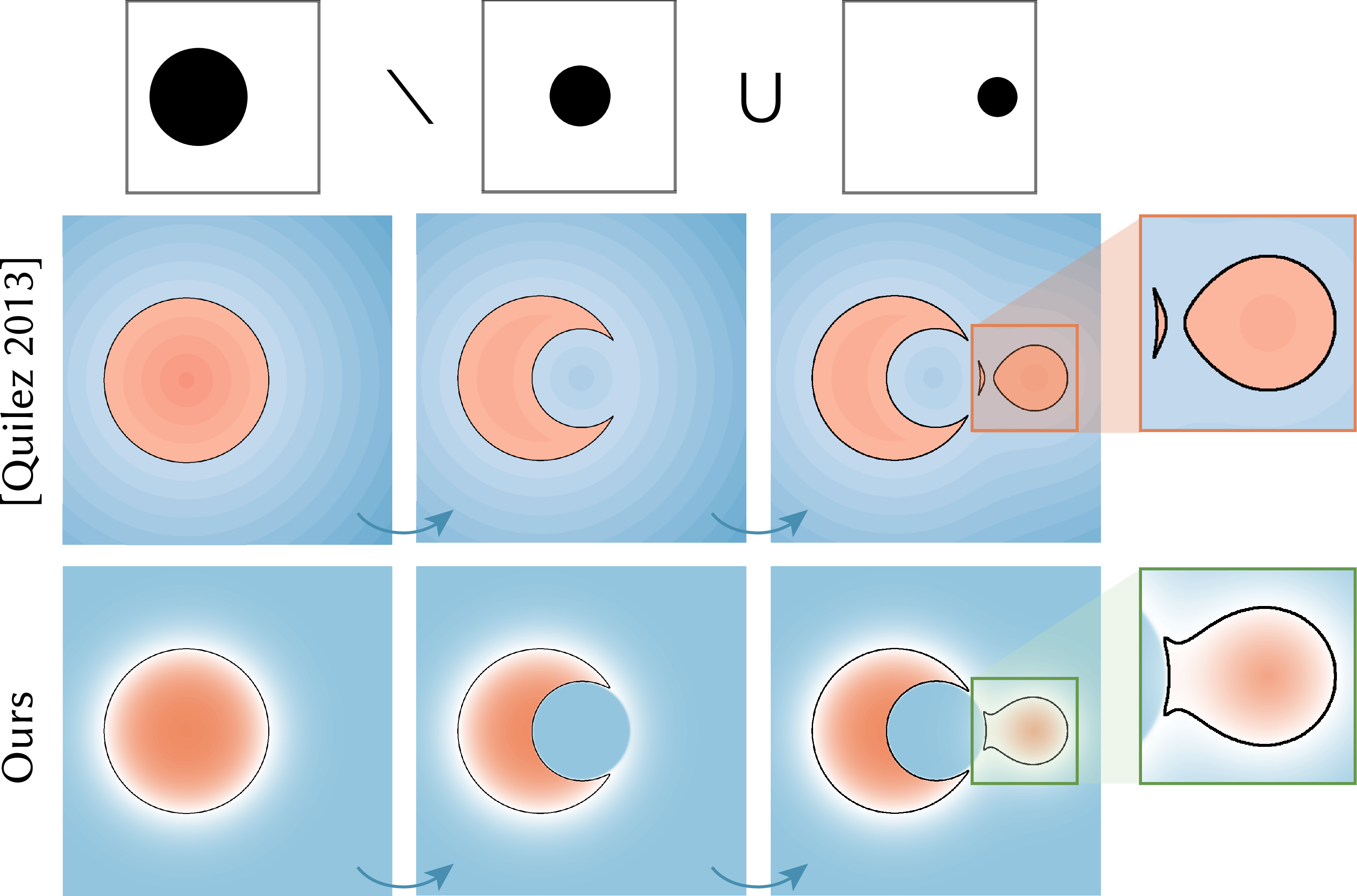}
    \end{center}
    \caption{Using the soft union presented by \citet{InigoSmoothMin} to compute the boolean expression leads to ``floating island'' artifacts (top row). This is because boolean operations on the signed distance function do not output a \emph{correct} signed distance function (see \cite{MarschnerSLJ23}). Our boolean operator operates on the occupancy function and remains an occupancy function after boolean operations. This leads to soft blending results that are free from artifacts (bottom row).}
    \label{fig:soft_boolean_pseudo_sdfs}
\end{figure}

This capability allows us to obtain visually indistinguishable results compared to the popular smoothed \min/\max{} \cite{InigoSmoothMin} operations on the signed distance function \edit{(\reffig{soft_boolean})}. 
Moreover, our approach is free from artifacts \edit{caused by discrepancy between the input and the output (see \reffig{soft_boolean_pseudo_sdfs}). This is because our method is \emph{closed}: both the input and the output are guaranteed to be soft occupancy functions.}
%
This is different from the previous method by \citet{InigoSmoothMin} such that their outputs are not signed distance functions \cite{MarschnerSLJ23}, even though the input is. 

\begin{figure}
    \begin{center}
    \includegraphics[width=1\linewidth]{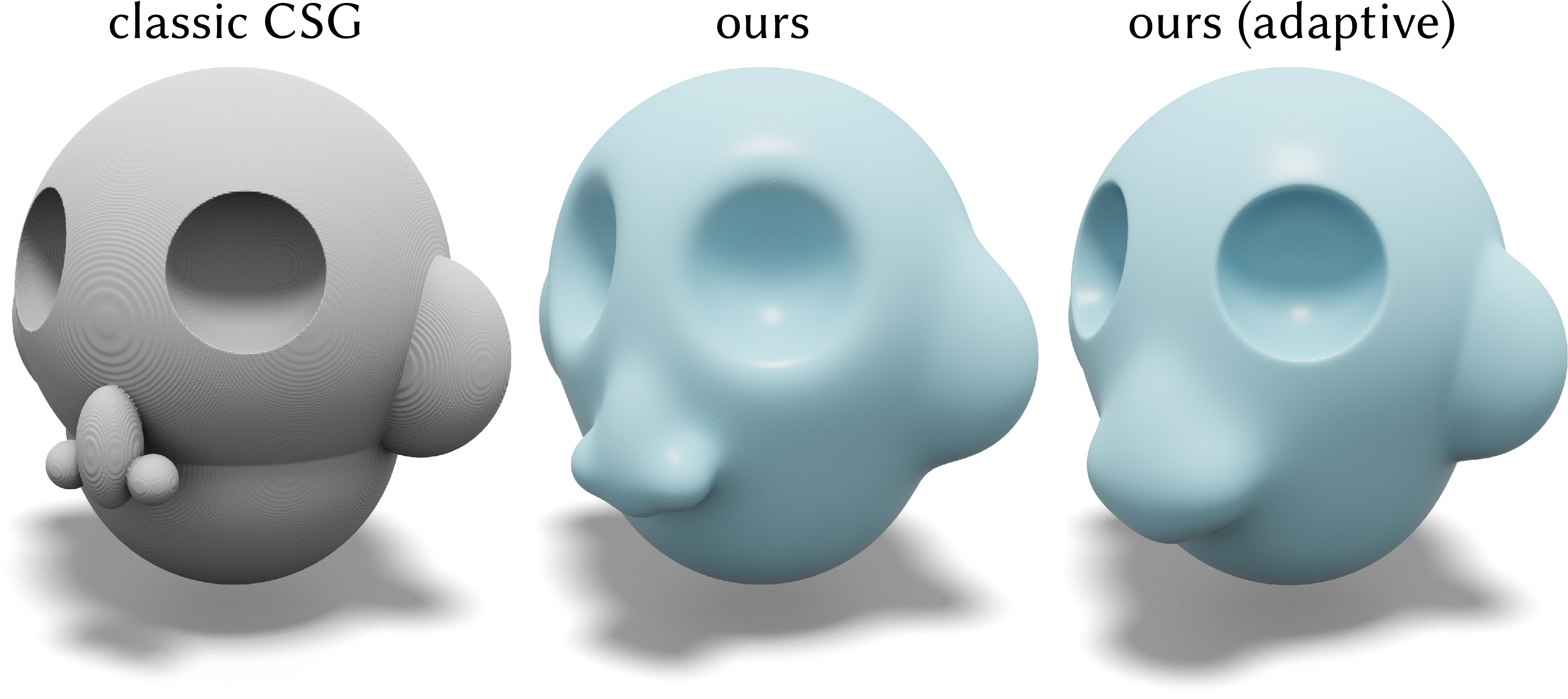}
    \end{center}
    \caption{Unlike classic CSG which can only model hard booleans (left), our method enables both crisp and smooth outputs (middle) and can control the smoothness adaptively (right) at the primitive level.}
    \label{fig:smooth_boolean_modeling}
\end{figure}

As the smoothness is controlled at the primitive level, we can easily have adaptive smoothness across the shape by simply changing the softness of each primitive occupancy (see \reffig{smooth_boolean_modeling}). Specifically, we consider primitive shapes represented as the signed distance function $s$, and we convert it to occupancy with the sigmoid function $\textit{sigmoid}(t\cdot s)$ with different softnesses by adjusting the positive temperature parameter $t$. 

\subsection{Single Shape Inverse CSG with Gradient Descent}\label{sec:inverse_csg_GD}
\begin{wrapfigure}[6]{r}{1.4in}
    \vspace{-12pt}
	\includegraphics[width=\linewidth, trim={50mm 0mm 0mm 0mm}]{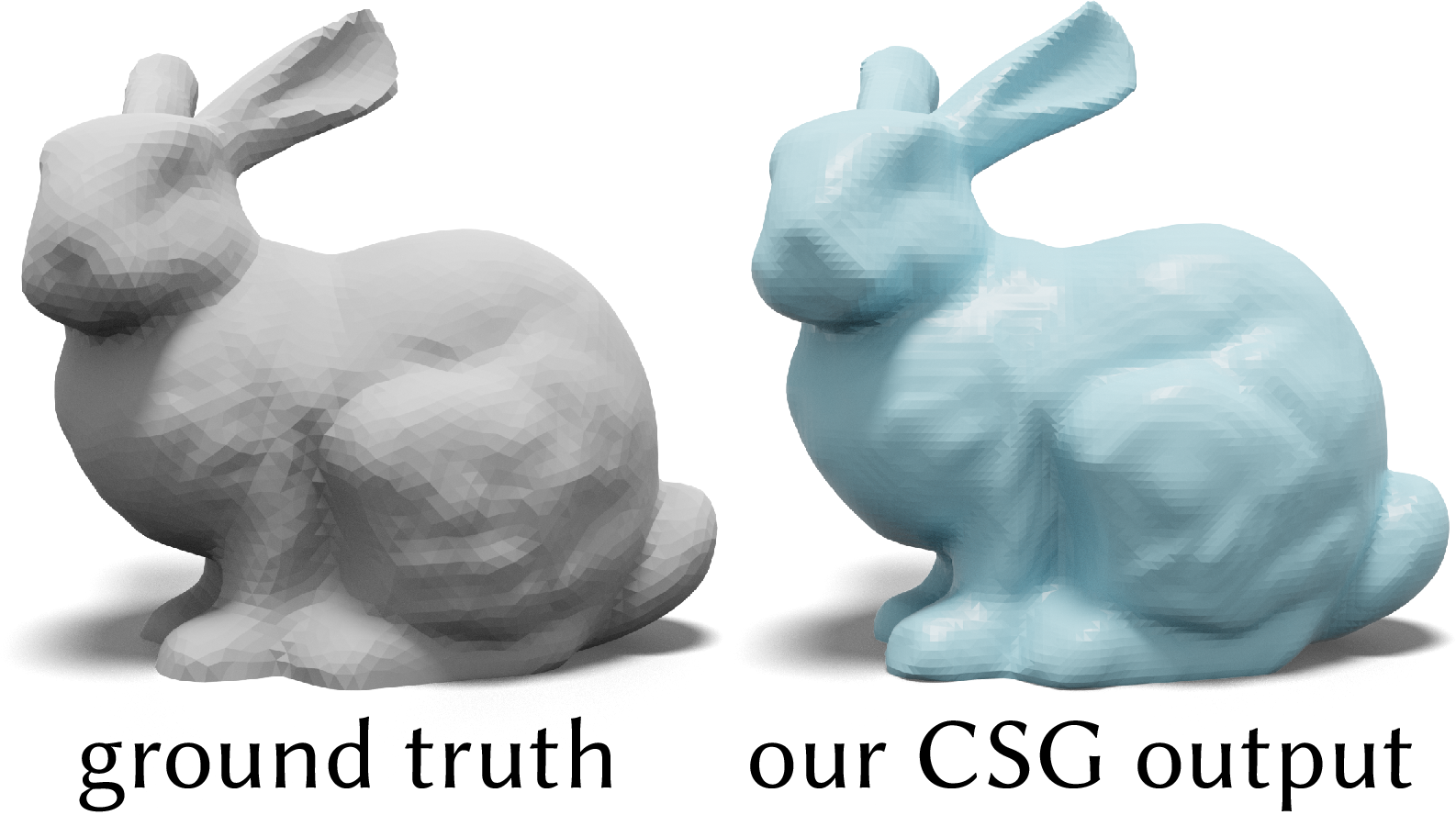}
	\label{fig:bunny_overfit}
\end{wrapfigure}
\begin{figure}
    \begin{center}
    \includegraphics[width=1\linewidth]{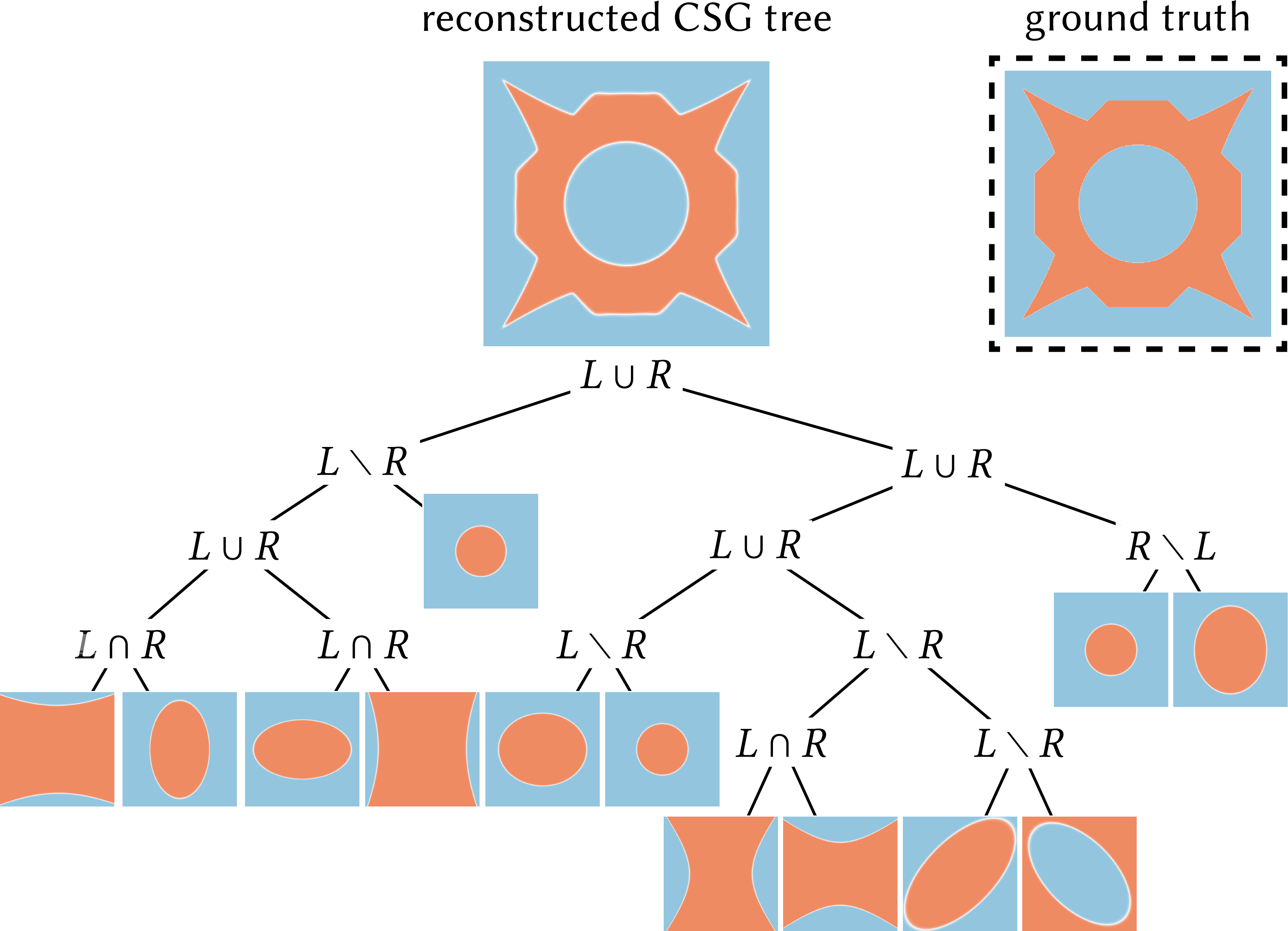}
    \end{center}
    \caption{After fitting, we run the classic CSG tree pruning \cite{tilove_csg_deletion} to reduce the size of the tree \edit{from 128 primitives initially down to 13.}}
    \label{fig:pruning}
\end{figure}
\begin{figure}
    \begin{center}
    \includegraphics[width=1\linewidth]{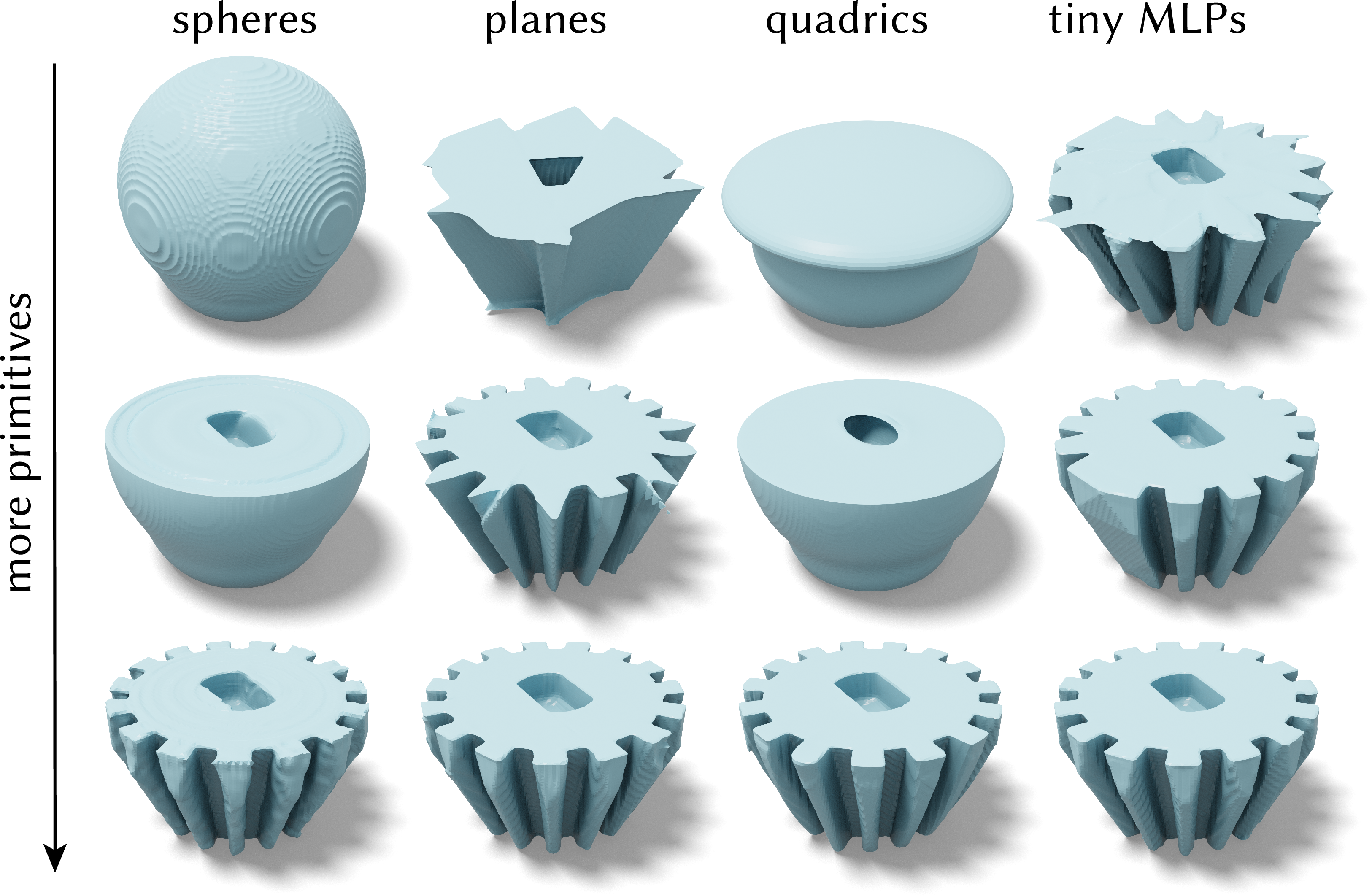}
    \end{center}
    \caption{Our method is applicable to different types of primitives, including spheres, planes, quadrics, and tiny neural implicit networks. The choice of primitives will change the inductive bias of the optimization, leading to favoring different results when the amount of primitives is insufficient.}
    \label{fig:one_shape_overfit}
\end{figure}
\begin{figure}
    \begin{center}
    \includegraphics[width=1\linewidth]{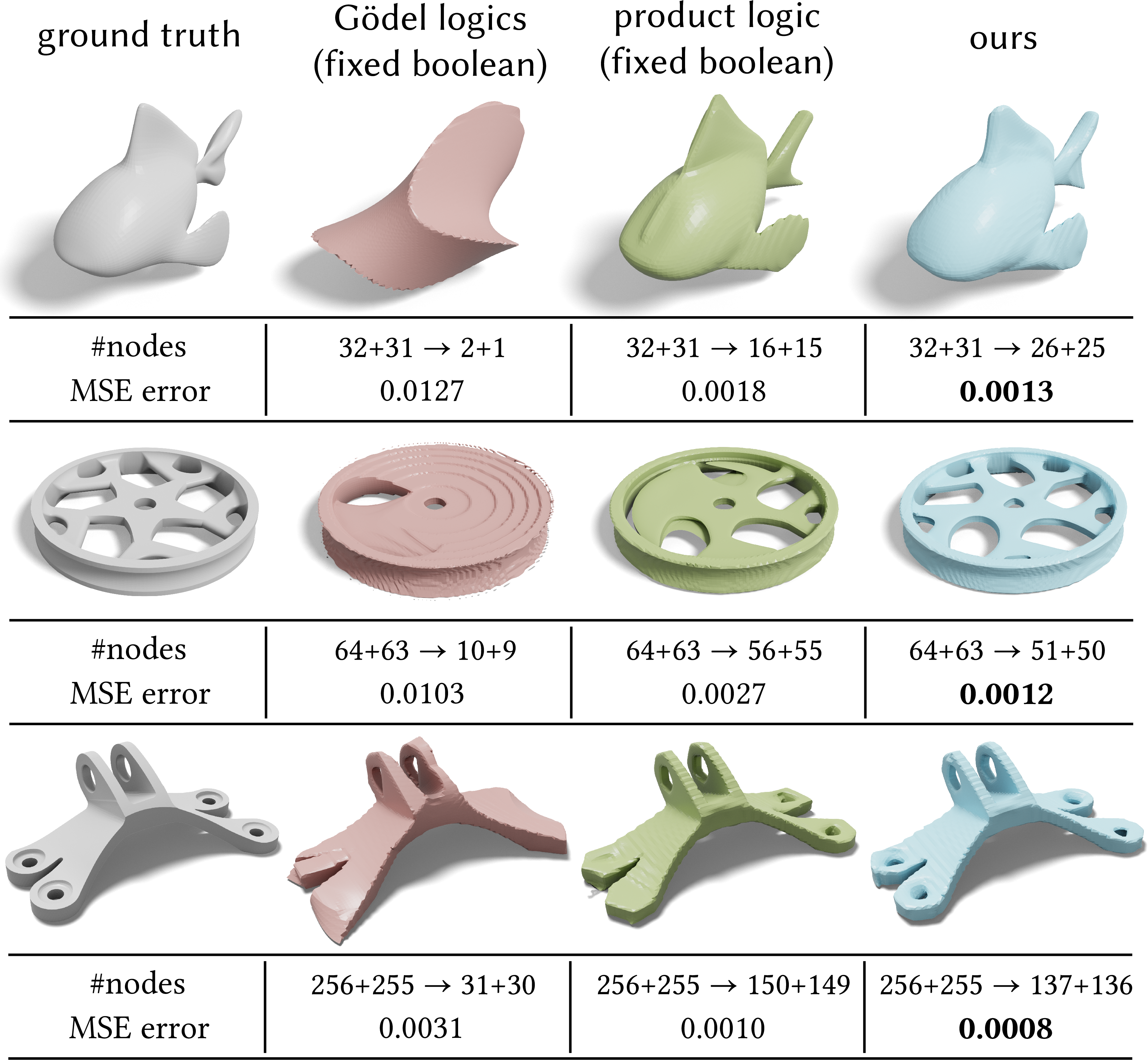}
    \end{center}
    \caption{Our method (blue) allows optimization of the type of boolean operations. This leads to a better fitting result compared to the product logic (green) and the G\"{o}del logic (red) with (randomly initialized) fixed boolean operations. We present the total number of nodes (primitive + boolean) before and after the optimization with pruning, showcasing improvements over different initial tree complexities. \edit{We also show the mean squared error on the occupancy evaluated on 2 million points sampled uniformly.}}
    \label{fig:ablation}
\end{figure}

Our approach enables us to simply use gradient descent to optimize a CSG tree that outputs a given shape (see \reffig{pruning}), even for smooth organic objects (see inset). 
Our method starts with a \emph{full} binary CSG tree (each boolean node has exactly two children, each primitive node is a leaf node) with randomly initialized (unified) boolean operators $\bool_\vc$ and primitive parameters (see \reffig{teaser}). Given a ground truth occupancy function, we minimize the \edit{mean square error} between the output occupancy from the CSG tree and the \edit{ground truth} with the ADAM optimizer \cite{KingmaB14}. To enforce our unified boolean operators converge to one of the boolean operations, we use a temperetured softmax function (see \refapp{inverse_csg_details} for implementation details).
At the end of the optimization, \edit{our result is} a set of optimized boolean nodes and primitive parameters. We then prune the redundant nodes with the classic CSG pruning \cite{tilove_csg_deletion} to obtain a more compact full binary CSG tree (see \reffig{pruning} \edit{and \refapp{inverse_csg_details} for implementation details}).
Our approach is independent of the choice of primitive families. We are able to convert a shape into CSG of spheres, planes, quadrics, or even tiny neural networks (\reffig{one_shape_overfit}).
%
%
Compared to fixing boolean nodes and only optimizing the primitive parameters, using our unified operator leads to better reconstruction (see \reffig{ablation}). 
%

\subsection{CSG Generative Models}\label{sec:csg_generation}
\begin{figure}
    \begin{center}
    \includegraphics[width=1\linewidth]{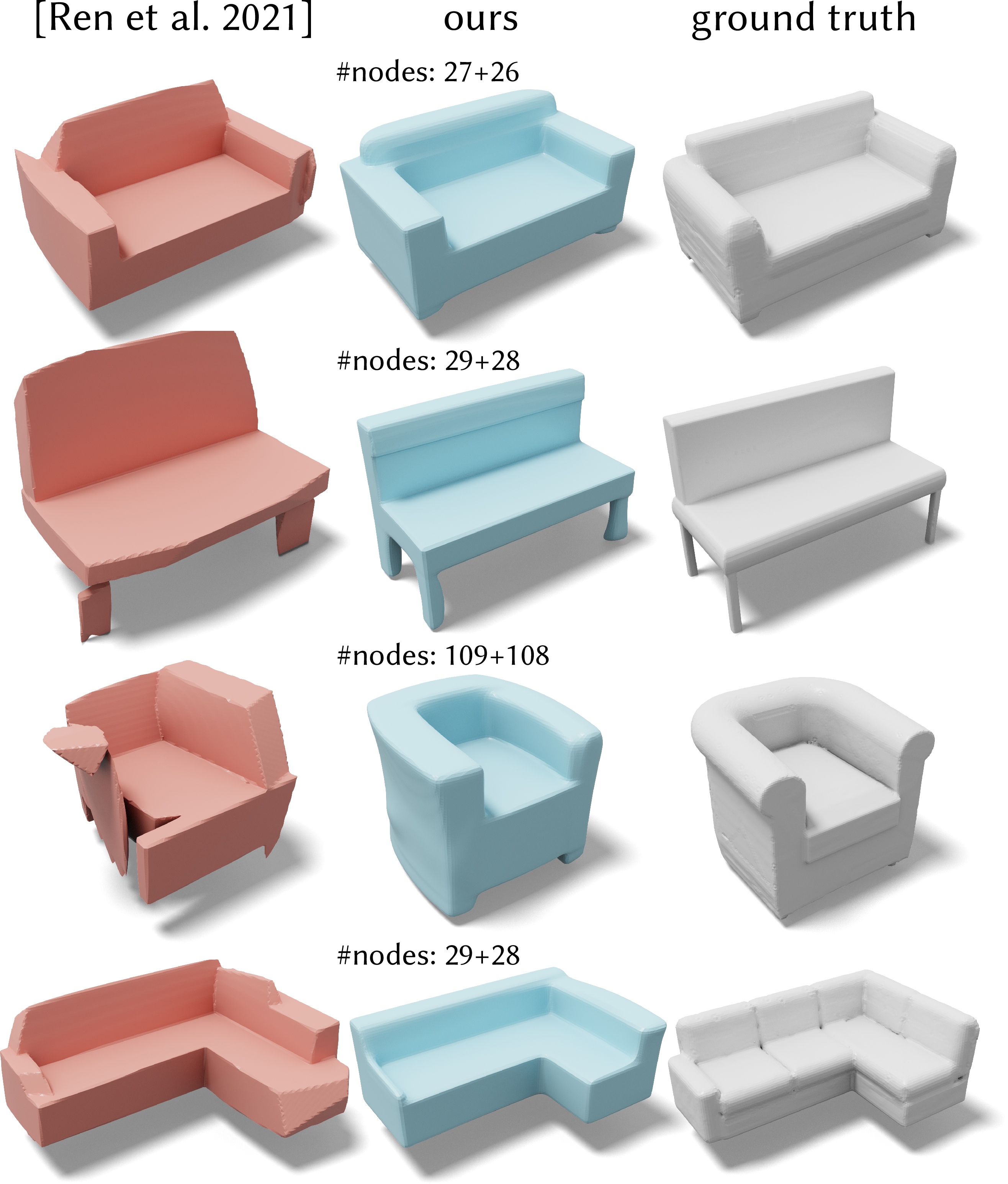}
    \end{center}
    \caption{By simply replacing the CSG tree structure in \cite{RenZ0LJCZPZZY21} with our method, our approach can obtain qualitative improvement in fitting the ground truth shape over their approach. We also report the number of nodes (\#primitives + \#boolean) after pruning for our method. Note that the raw output of \cite{RenZ0LJCZPZZY21} requires around 15K boolean operations with many redundant ones. Thus, we do not report their number because a pruning method for them would be required for a fair comparison on the compactness of the tree.  }
    \label{fig:csg_stumps}
\end{figure}

We demonstrate how our method can improve existing methods for fitting a shape dataset and generating CSG trees. 
Specifically, we focus on improving a hypernetwork approach proposed by \citet{RenZ0LJCZPZZY21}. In short, given a point cloud, they propose to train a hypernetwork conditioned on the point cloud to output the parameters of their proposed CSG tree structure with pre-determined boolean operations. 
To demonstrate improvements over their method, we conduct the experiment on the same dataset, loss function, and hypernetwork, but we replace their CSG tree structure with our fuzzy boolean CSG tree. 
With such change, we demonstrate noticeable improvement over both qualitative \reffig{csg_stumps} and quantitative \reftab{quantative_csg_stumps} evaluations. 
The baseline \cite{RenZ0LJCZPZZY21} is based on their implementation and their pre-trained model weights.
\begin{table}[t]
    \setlength{\tabcolsep}{5.425pt}
    \centering
    \caption{By swapping the decoder in \cite{RenZ0LJCZPZZY21} with a decoder based on our unified boolean operator, we achieve improvements in quantitative metrics, including the mean squared error (MSE), classification accuracy, and F-score, on the ShapeNet dataset \cite{ChangFGHHLSSSSX15}. We provide the maximum number of nodes in the decoder (\#primitives + \#boolean nodes) and show that, despite being more compact, our approach still leads to better reconstruction.}
    \begin{tabularx}{0.9\linewidth}{l|ccc|c}
        \toprule
        \text{} & \textit{MSE} & \textit{Accu.} & \textit{F-score} & \textit{total \#nodes} \\
        \midrule
        \text{Ren et al. 2021}  & 0.049 & 0.912 & 0.938 & 256+65792 \\
        \text{Ours}  & \textbf{0.018} & \textbf{0.982} & \textbf{0.978} & \textbf{512+511} \\
        \bottomrule
    \end{tabularx}
    \smallskip
    \label{tab:quantative_csg_stumps}
\end{table}
\section{Limitations \& Future Work}

We introduce a unified differentiable boolean operator for solid shapes with soft occupancy. Our approach enables optimization of both the primitives and the boolean operations with continuous optimization techniques, such as gradient descent. As a preliminary investigation, the efforts described here open a cast of new directions for future work as well as room for improvement.  We describe next a set of the limitations and opportunities for additional next steps and research directions.

\paragraph{Optimizing Tree Structures}
Although we have enabled differentiation through boolean and primitive nodes, currently, the structure of our tree is held fixed during optimization. 
Despite fitting the shape well, our approach often leads to complicated CSG trees as a result, even after pruning. 
We believe future research in optimizing among tree structures and identifying when to grow/prune/rotate the tree nodes would be beneficial to reduce tree complexity. 

\paragraph{Tree Properties}
The ability to optimize the tree structure could unlock optimizing the tree to have certain properties, such as compactness or editability. A well-known challenge of inverse CSG is that a shape can be constructed by an infinite number of different CSG trees. We suffer from the same issue that our approach only finds one of the trees, but there is no guarantee that the tree we obtain is, for instance, the most compact option. 

\paragraph{Extended Fuzzy CSG System}
In our work, we explore how fuzzy logic may be applied to CSG modeling. We evaluate the fuzzy counterparts of existing CSG operations (\union, \intersection, \textsc{difference}), but there are more fuzzy logic operators that do not exist in CSG traditionally. For instance, the \emph{fuzzy aggregation} operator \cite{fuzzy_logic_book} can be perceived as a generalization of {\union} or {\intersection} on a collection of primitive shapes, instead of two. Adding such operators could enable new possibilities in tree structure optimization by, for instance, selecting which primitives to use when performing boolean operations.

\paragraph{Hardware Acceleration}
Our current fuzzy CSG system is based on an un-optimized implementation of fuzzy logic operators. However, as shown in several other fields, fuzzy logic operators can be greatly accelerated with parallel hardware implementations (e.g., \cite{ontiveros2016hardware}). A hardware-accelerated version of our CSG system based on fuzzy logic could accelerate our method to run in real time.

\paragraph{CSG Generative Models}
Making CSG systems differentiable could be beneficial for future exploration on (black boxed) neural symbolic generative models \cite{RitchieGJMSWW23} that output (white boxed) CSG tree parameters. 
As this is orthogonal to our contributions, we simply evaluate our method based on the off-the-shelf architecture in \refsec{csg_generation}. Future work on better neural network architectures for tree generation would be beneficial to empower CSG generation.  
\\

\noindent 
CSG modeling with optimization offers alternatives to mesh-based geometry representations that can be compact and less resolution dependent.  Our exploration of the \edit{fuzzy} boolean operator brings the automatic production of CSG models one step closer and opens avenues for further advances including speed-ups and relaxed constraints on CSG hierarchies.
\edit{Inspired by the connections between fuzzy logic and other graphics applications (e.g., image/volumetric compositing), exploring applications of fuzzy logic beyond CSG could be an interesting future direction.}

\bibliographystyle{ACM-Reference-Format}
\bibliography{sections/references}
\clearpage
\appendix

\section{Implementation Details for Inverse CSG}\label{app:inverse_csg_details}
\paragraph{Initialization} 
Our method starts with a randomly initialized full binary CSG tree that consists of our fuzzy boolean nodes \refequ{unified_boolean} and primitive shape represented as soft occupancy functions.
\edit{We initialize the parameters of the boolean and primitive nodes with a uniform distribution between -0.5 and 0.5.}
As the required tree complexity is unknown, we initialize a ``big'' CSG tree (e.g., 1024 primitive shapes) to reduce the chance of having an insufficient number of primitives. 

\paragraph{Primitive Choices} 
In terms of the choice of primitives, except the one in \reffig{one_shape_overfit}, we use quadric surfaces $q$
\begin{align}
    q(x,y,z) =&\; q_0 x^2 + q_1 y^2 + q_2 z^2 \\
    &+ q_3 xy + q_4 yz + q_5 zx + q6 x + q_7 y + q_8 z + q_9 
\end{align}
in all our experiments partly due to its popularity in industry \cite{samuel1976methodology}.
More crucially, we believe using a less expressive primitive (compared to MLPs) give us a clearer signal on the performance of our proposed boolean operator. This is because an expressive primitive family, such as a big neural network, is able to fit a shape even without using any boolean operations.
Then we convert the quadric function into a soft occupancy function with the \textit{sigmoid} function
\begin{align}
    o(x,y,z) = \textit{sigmoid}(s \times q(x,y,z))
\end{align}\
where $s$ is a trainable ``sharpness'' parameter to uniformly scale the quadric function to make it sharper or smoother. This allows the model to change the sharpness of quadric surface without changing the shape. 
Empirically, we notice a better convergence rate with a trainable sharpness. 

\paragraph{Boolean Parameterization} 
The side effect of having a unified boolean operator in is the possibility of not converging the one of the boolean operations. We alleviate this issue by parameterizing $\vc$ with $\tilde{\vc} \in \R^4$ as
\begin{align}
    \vc = \textit{softmax}(\sin(\omega \tilde{\vc}) \cdot t)
\end{align}
where $t \in \R$ is the temperature. 
We leverage the \textit{softmax} function to ensure the resulting $\vc$ is always a valid barycentric coordinate. We set the temperature $t$ to a high value (e.g., $t = 10^3$) to encourage $\vc$ to be numerically close to a one-hot vector for most parameter choices of $\tilde{\vc}$. The $\sin(\omega \cdot)$ (with $\omega = 10$) function is to ensure boolean operator type can still be changed easily in the later stage of the optimization. Without it, changing $\vc$ will require many iterations when $\tilde{\vc}$ has a large magnitude because each gradient update only updates $\tilde{\vc}$ a little. We observe that this parameterization of $\vc$ converges to a one-hot vector in all our experiments, even though we only softly encourage most parameter choices of $\tilde{\vc}$ to be one-hot vectors. 
We suspect this is because any in-between operations will have occupancy values away from 0 or 1, whereas the target shape has binary occupancy values, converging to in-between operations can still occur when imperfect fitting happens.

\paragraph{Optimization} 
We define the loss function as the mean square error between the output occupancy from the CSG tree and the ground truth occupancy, evaluated on some sampled 3D points.
We sample the points with approximately 40\% on the surface, 40\% near the surface, and 20\% randomly in the volume. 
We regenerate these sampled point every couple iterations (e.g., 10) to make sure we sample most areas in the volume.
We use the ADAM optimizer \cite{KingmaB14} with learning rate 1e-3 to train our model. 

\paragraph{Pruning} 
After training, we prune redundant primitive/boolean nodes with post-processing. 
%
\edit{To determine redundant nodes, we follow the definition proposed by \citet{tilove_csg_deletion} to characterize each boolean or primitive node as either \emph{W-redundant} or \emph{$\varnothing$-redundant}. Intuitively,} given a boolean node and its two child subtrees, if a subtree can be replaced with a full (soft occupancy with all 1s) or an empty (soft occupancy with all 0s) function without changing the output after the boolean operation, then this node is redundant and can be removed. 
%
%
We generalize such a redundancy definition to fuzzy boolean operations by setting a small threshold (e.g., mean squared soft occupancy error $10^{-3}$) to determine whether the difference after replacing a subtree with full/empty function is small enough. 
\edit{With the notion of redundancy, we visit each node in the CSG tree in \emph{post-order} and delete the node (including its children) if it is classified as a redundant node.}
We demonstrate the effectiveness of such a simple pruning strategy to greatly reduce the complexity of the optimized CSG tree in \reffig{pruning}.

\edit{
\paragraph{Linear Time CSG Forward Pass}
During training, the full binary tree structure can be implemented in parallel for each layer by leveraging the fact that the number of node is pre-determined.
However, after pruning, the tree structure becomes irregular.
Running the forward pass after pruning requires graph traversal from the leaf primitive nodes to the boolean nodes and all the way to the root boolean node. 
To facilitate efficient inference, we employ the linear time traversal algorithm proposed by \citet{GrasbergerDWLR16} to speed up the forward pass. 
Their key idea is to traverse the CSG tree in post-order and push/pop intermediate results from a \emph{stack}. This traversal has a continuous memory storage of all the nodes and only requires reading each node once. 
}

\end{document}